\documentclass[aps,prb,twocolumn,superscriptaddress]{revtex4-1}
\usepackage{graphicx}
\usepackage{ulem}
\usepackage{color}\usepackage{xspace}\usepackage{amsmath}
\usepackage{siunitx}
\usepackage{miller}

\usepackage[version=3]{mhchem}
\usepackage{natbib}
\usepackage[]{hyperref}
\usepackage{bm}

\renewcommand{\deg}{$^{\circ}$\,}

\newcommand{\tbmno}{TbMnO$_3$ }

\newcommand{\rmnokomma}{$R$MnO$_3$}

\newcommand{\mnwo}{MnWO$_4$ }
\newcommand{\mnwokomma}{MnWO$_4$}

\newcommand{\nafe}{NaFe$(\text{WO}_4)_2$ }
\newcommand{\nafekomma}{NaFe$(\text{WO}_4)_2$}

\begin{document}


\title{Structural dimerization in the commensurate magnetic phases of \nafe and \mnwo}

\author{S. Biesenkamp}
\affiliation{$I\hspace{-.1em}I$. Physikalisches Institut,
Universit\"at zu K\"oln, Z\"ulpicher Str. 77, D-50937 K\"oln,
Germany}

\author{N. Qureshi}
\affiliation{Institut Laue-Langevin, 71 avenue des Martyrs, CS 20156, 38042 Grenoble Cedex 9, France }

\author{Y. Sidis}
\affiliation{Laboratoire L\'eon Brillouin, C.E.A./C.N.R.S., F-91191 Gif-sur-Yvette CEDEX, France}

\author{P. Becker}
\affiliation{Abteilung  Kristallographie,  Institut  f\"ur  Geologie  und  Mineralogie, Universit\"at zu K\"oln, Z\"ulpicher Str.  49b,  50674  K\"oln,  Germany}

\author{L. Bohat\'{y}}
\affiliation{Abteilung  Kristallographie,  Institut  f\"ur  Geologie  und  Mineralogie, Universit\"at zu K\"oln, Z\"ulpicher Str.  49b,  50674  K\"oln,  Germany}

\author{M. Braden}\email[e-mail: ]{braden@ph2.uni-koeln.de}
\affiliation{$I\hspace{-.1em}I$. Physikalisches Institut,
Universit\"at zu K\"oln, Z\"ulpicher Str. 77, D-50937 K\"oln,
Germany}





\date{\today}

\begin{abstract}
The structural distortion and magnetoelastic coupling induced through commensurate magnetism has been investigated by neutron diffraction in structurally related MnWO$_4$ and NaFe$(\text{WO}_4)_2$.
Both systems exhibit a competition of incommensurate spiral and commensurate spin up-up-down-down ordering along the magnetic chains.
In the latter commensurate phases, the alternatingly parallel and antiparallel arrangement of Fe$^{3+}$ respectively Mn$^{2+}$ moments leads to sizeable bond-angle modulation and thus to magnetic dimerization.
For NaFe$(\text{WO}_4)_2$ this structural distortion has been determined to be strongest for the low-field up-up-down-down arrangement, and the structural refinement yields a bond-angle modulation of $\pm 1.15(16)$ degrees.
In the commensurate phase of MnWO$_4$, superstructure reflections signal a comparable structural dimerization and thus strong magneto-elastic coupling different to that driving the multiferroic order.
Pronounced anharmonic second- and third-order reflections in the incommensurate and multiferroic phase of MnWO$_4$ result from tiny commensurate fractions that can depin multiferroic domains.

\end{abstract}


\maketitle


\section{\label{sec:level1}Introduction}
Since the discoveries of giant magnetoelectric coupling in TbMnO$_3$ \cite{Kimura2003} and of functional responses in thin
film BiFeO$_3$ \cite{Wang2003}, the research interest on multiferroic materials has been tremendously increased \cite{Fiebig_2005,Khomskii2009,Fiebig2016,Spaldin2019}. In type-II multiferroics a ferroic arrangement directly induces a second ferroic order, which implies a strong coupling and thus the potential of manipulating one ferroic parameter by the conjugate field of the coexisting order \cite{Fiebig_2005,Khomskii2009,Fiebig2016}.

In many type-II multiferroics the microscopic coupling mechanism is described by the inverse Dzyaloshinskii-Moriya interaction (DMI) \cite{Dzyaloshinsky1958,Moriya1956}
relating the ferroelectric polarization to the cross product of  two spins and the direction of their connecting vector $\bm{e_{ij}}$: $\bm{P}\propto\bm{e_{ij}}\times(\bm{S}_i\times\bm{S}_j)$ \cite{Katsura2005,Mostovoy2006,Sergienko2006a,Kimura2007}.
However, due to its origin from spin-orbit coupling, this effect and especially the induced ferroelectric polarization remain small compared to conventional ferroelectrics \cite{Khomskii2009}. In contrast, larger ferroelectric polarization and more generally larger magnetoelastic coupling can be generated by symmetric-exchange striction in a collinear magnetic structure \cite{Sergienko2006}.
Indeed, exchange striction was observed to induce large ferroelectric polarization in rare-earth manganates  \rmnokomma{,} where a small $R$ ionic radius causes a change of the
magnetic structure from an incommensurate spiral to commensurate E-type spin up-up-down-down ($uudd$) arrangement \cite{Sergienko2006,Ishiwata2010,Kimura2005}.

The same kind of magnetic competition between incommensurate and commensurate magnetic order can also be found in the phase diagrams of the two tungstates
\mnwo and \nafekomma \cite{Ehrenberg1997a,Heyer2006,Lautenschlager1993,Holbein2016}. But only in \mnwo the incommensurate phase is multiferroic arising from the inverse DMI. In contrast, both materials exhibit strong magnetoelastic coupling in their commensurate phases with $uudd$ spin arrangement.
Characterizing the magnetoelastic coupling associated with the commensurate $uudd$ order in these tungstates is the aim of this work.

The crystal and magnetic structures are illustrated in Fig.~1. Both materials are characterized by
zigzag chains of edge-sharing MnO$_6$- respectively FeO$_6$ octahedra along $c$ that are adjacent along $b$. Along the third direction, $a$, the magnetic chains
are separated by layers of WO$_6$ octahedra in MnWO$_4$. This separation is enhanced in \nafekomma , whose structure results from that of MnWO$_4$ by
substituting the Mn in the zigzag chain layers alternatingly with Fe and non magnetic Na. Therefore, every second layer is magnetic in MnWO$_4$, while only every
fourth layer is magnetic in \nafekomma . Concerning the magnetic structures it is most important to analyze the intrachain arrangements depicted in Fig.~1.
MnWO$_4$ undergoes first a transition into a collinear incommensurate phase AF3 and then into an incommensurate spiral phase, AF2. This AF2 phase can be compared to the low-field
incommensurate phase (LF-IC) that \nafe directly reaches upon cooling. While the AF2 phase in MnWO$_4$ is multiferroic with finite ferroelectric polarization, the
LF-IC phase remains paraelectric in \nafekomma . The difference between the two magnetic structures concerns the rotation of the magnetic moments in the cycloid plane \cite{Ehrenberg1997a,Lautenschlager1993,Holbein2016}. The upper and lower
magnetic moments in one zigzag chain rotate in the same sense in MnWO$_4$ and in opposite sense in \nafekomma . Therefore, the contributions to the polarization according
to the inverse DMI cancel out for the Fe compound. MnWO$_4$ reaches a commensurate phase, AF1, at low temperature characterized by $uudd$ order within a chain \cite{Ehrenberg1997a,Lautenschlager1993}, and also
\nafe exhibits commensurate phases, low-field commensurate (LF-C) and high-field commensurate (HF-C), that however differ for high and low magnetic fields. In addition there is a strong hysteresis so that the magnetic structure
depends on how a point in the $B$,$T$ phase diagram (see inset of figure \ref{sweep_1}) is reached \cite{Holbein2016}.
In contrast to the E-type manganates, the $uudd$ structures occurring in MnWO$_4$ and in  \nafe are not multiferroic as there is no finite polarization. Nevertheless,
we will show that also these $uudd$ phases show a strong magnetoelastic coupling characterized by prominent  structural dimerization. Note, that the ferroelectric polarization in \rmnokomma \
only arises from the interplay between bond-angle driven dimerization and octahedron rotation, which causes a shift of all bridging oxygen ions in the same direction \cite{Sergienko2006,Ishiwata2010}.

The first-order transition to the commensurate AF1 phase in \mnwo is accompanied by sizable thermal expansion anomalies ($\frac{\Delta b}{b}\approx -1\times 10^{-5}$) \cite{Chaudhury2008}. 
Already in the multiferroic AF2 phase a considerable magneto-elastic coupling was observed \cite{Finger2010,Finger2010a} in form of magnetic and nuclear anharmonic modulations.
Every incommensurate modulated magnetic structure enforces a nuclear modulation of half the period in real space and of twice the propagation vector. Indeed a strong second order reflection appears in the multiferroic phase \cite{Taniguchi2008}, and neutron polarization analysis separates the magnetic and nuclear contribution revealing a strong nuclear fraction and a small magnetic part \cite{Finger2010}.
The same competition of incommensurate and $uudd$ ordering also appears in the double tungstate \nafekomma{,} in which transitions to the commensurate phases are accompanied by thermal expansion anomalies that are at least one order of magnitude larger ($\frac{\Delta b}{b}\approx -2.6\times 10^{-4}$) compared to the observed anomalies in \mnwo \cite{Holbein2016,Chaudhury2008}. Independently of the absence of multiferroicity in \nafekomma{,} the similar competition of incommensurate spiral and $uudd$ ordering as well as the very strong magneto-elastic coupling to the commensurate phases render \nafe an ideal candidate to investigate the magneto-elastic coupling in form of a structural dimerization effect, when entering the commensurate $uudd$ phases.

\begin{figure}
 \includegraphics[width=\columnwidth]{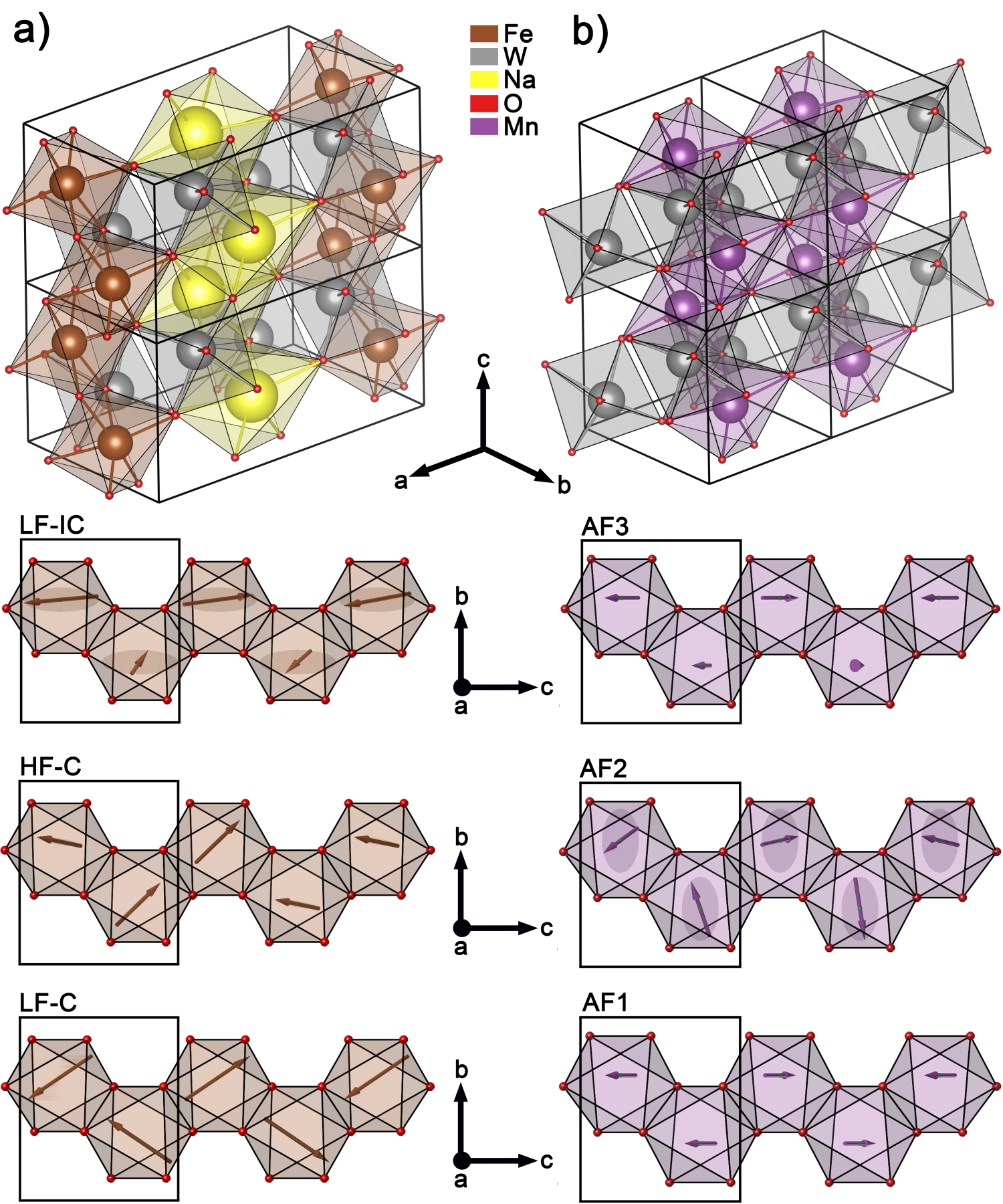}
  \caption{\label{structure}(a) and (b) display the nuclear and magnetic structure of \nafe and \mnwokomma{,} respectively. The unit cells are marked by black boxes and arrows indicate the direction of magnetic moments within the respective magnetic phases. Note the similarities between the incommensurate spiral phases AF2 and LF-IC, as well as between the commensurate $uudd$ phases AF1 and HF-C / LF-C.}
  \label{disp}
 \end{figure}
 
The research field of multiferroics not only concerns fundamental aspects but also the development of new sensor and memory devices utilizing multiferroicity \cite{Scott2007a,Bibes2008,Zhai2006a}.
For memory devices the writing speed is an important factor, which in the case of a multiferroic material relates to the relaxation and inversion processes of antiferromagnetic domains, that are controlled by electric fields. The electric-field dependent behavior of  multiferroic domains was studied intensively for the prototype multiferroic materials \tbmno and \mnwo \cite{Arkenbout2006,Meier2009,Niermann2014,Baum2014,Stein2015,Stein2017}. A time-resolved neutron scattering investigation as well as second-harmonic generation imaging of multiferroic domain inversion in \mnwo revealed a rather complex switching behavior, as the inversion process is speeding up towards the lower first-order transition to the commensurate $uudd$ phase\cite{Baum2014,Hoffmann2011a}.

In this paper we report studies of the structural dimerization  appearing in the commensurate $uudd$ phases in \nafe and in \mnwokomma{.}
The paper is organized as follows: After the experimental methods, we first discuss the results for \nafekomma{,} where a stronger dimerization is suggested by the larger thermal expansion anomalies \cite{Chaudhury2008,Holbein2016}. Besides a qualitative proof of structural dimerization, we present a structural refinement that quantifies the structural distortion in \nafekomma{.} In the last part of this paper we discuss the structural dimerization in \mnwo and its relation to the higher harmonic reflections in the incommensurate phase
as well as its role in depinning multiferroic domain walls.

  \section{\label{sec:exp_method}Experimental Methods}

The experiments on the structural dimerization in \nafekomma \ were performed on the D10 diffractometer at the Institut Laue-Langevin (ILL).
For the expected bond-angle modulation within the magnetic zigzag chain the shift of light oxygen ions is crucial, and therefore neutron diffraction is the appropriate tool. The effect is expected to be small and hence, it was essential to obtain the best possible signal to background ratio. Thus the instrument had to be equipped with the vertically focussing pyrolytic graphite analyzer, which reduces the background significantly, and a single He$^3$ detector was used. The commensurate phases of \nafe are only accessible by applying a magnetic field along $b$. Thus, the sample was mounted in the $(1\:0\:0)/(0\:0\:1)$ scattering plane and a vertical cryomagnet was used ($\mu_0H<\SI{6}{\tesla}$). All temperature and magnetic field dependent measurements have been performed with $\lambda=\SI{2.36}{\angstrom}$ (pyrolithic graphite monochromator) yielding high flux, and the data collection for a structural refinement was done with $\lambda=\SI{1.26}{\angstrom}$ (copper monochromator) in order to reach more reflections. The single crystal with dimensions $\SI{1}{\milli\meter}\times\SI{5}{\milli\meter}\times\SI{10}{\milli\meter}$ was grown as described in reference \onlinecite{Holbein2016}, which also presents the characterization
by various macroscopic methods.

The measurements on \mnwokomma were performed on the triple-axis spectrometer 4F2 at the Laboratoire L\'{e}on Brillouin (LLB)  using a wavelength $\lambda=\SI{4.05}{\angstrom}$
extracted with a highly-oriented pyrolithic graphite doublemonochromator. Higher order contaminations were suppressed by a cooled Be filter and the sample was cooled
in a closed-cycle refrigerator. The crystal had dimensions of $\SI{2}{\milli\meter}\times\SI{4}{\milli\meter}\times\SI{9}{\milli\meter}$ and was mounted in an aluminium can
containing He exchange gas. The scattering geometry was  $(1\:0\:0)/(0\:0\:1)$. Information concerning the crystal growth and characterization are given in references \onlinecite{Finger2010,Finger2010a,Stein2017}.

 \section{\label{sec:level1}Results and Discussion}
 \subsection{\label{sec:level2}Structural dimerization in \nafe}

 \nafe crystallizes in the $P2/c$ space group with $a\approx\SI{9.88}{\angstrom}$, $b\approx\SI{5.72}{\angstrom}$, $c\approx\SI{4.94}{\angstrom}$ and $\beta\approx\SI{90.33}{\degree}$ (at $\SI{298}{\kelvin}$) \cite{Klevtsov1970}.
 The crystallographic unit cell of \nafe is doubled along $a$ with respect to the unit cell of \mnwo (see Fig.~\ref{structure}(a)). Due to the enhanced distance between adjacent [FeO$_6$]-octahedra chains along $a$, the magnetic interaction is significantly reduced in this direction, which provokes a more two-dimensional nature of magnetic ordering in \nafekomma{.} Nevertheless, the system exhibits a cascade of magnetically ordered phases (see Fig.~\ref{structure}(a)) \cite{Holbein2016}. First, below $T_{N}\approx\SI{4}{\kelvin}$ an incommensurate three-dimensional magnetic order with $\bm{k}=(0.485,\frac{1}{2},0.48)$ develops. The moments are rotating in a plane, which is spanned by the easy plane $e_{ac}$ and a small $b$ component. However, as neighboring spirals rotate in the opposite sense with respect to each other, the DMI effect cancels and no multiferroic state develops in the low-field incommensurate (LF-IC*) phase. Below $T\approx\SI{3}{\kelvin}$ emerging third-order reflections signal a squaring up of the magnetic structure but the spiral magnetic structure remains incommensurate in zero field at low temperatures (LF-IC phase). When applying a magnetic field at low temperatures ($\mu_0H\approx \SI{2}{\tesla}$ at $T\approx \SI{2.5}{\kelvin}$), the system undergoes a spin-flip transition to a high-field commensurate $uudd$ arrangement with $\bm{k}=(\frac{1}{2},\frac{1}{2},\frac{1}{2})$ (HF-C phase).
 \begin{figure}
 \includegraphics[width=\columnwidth]{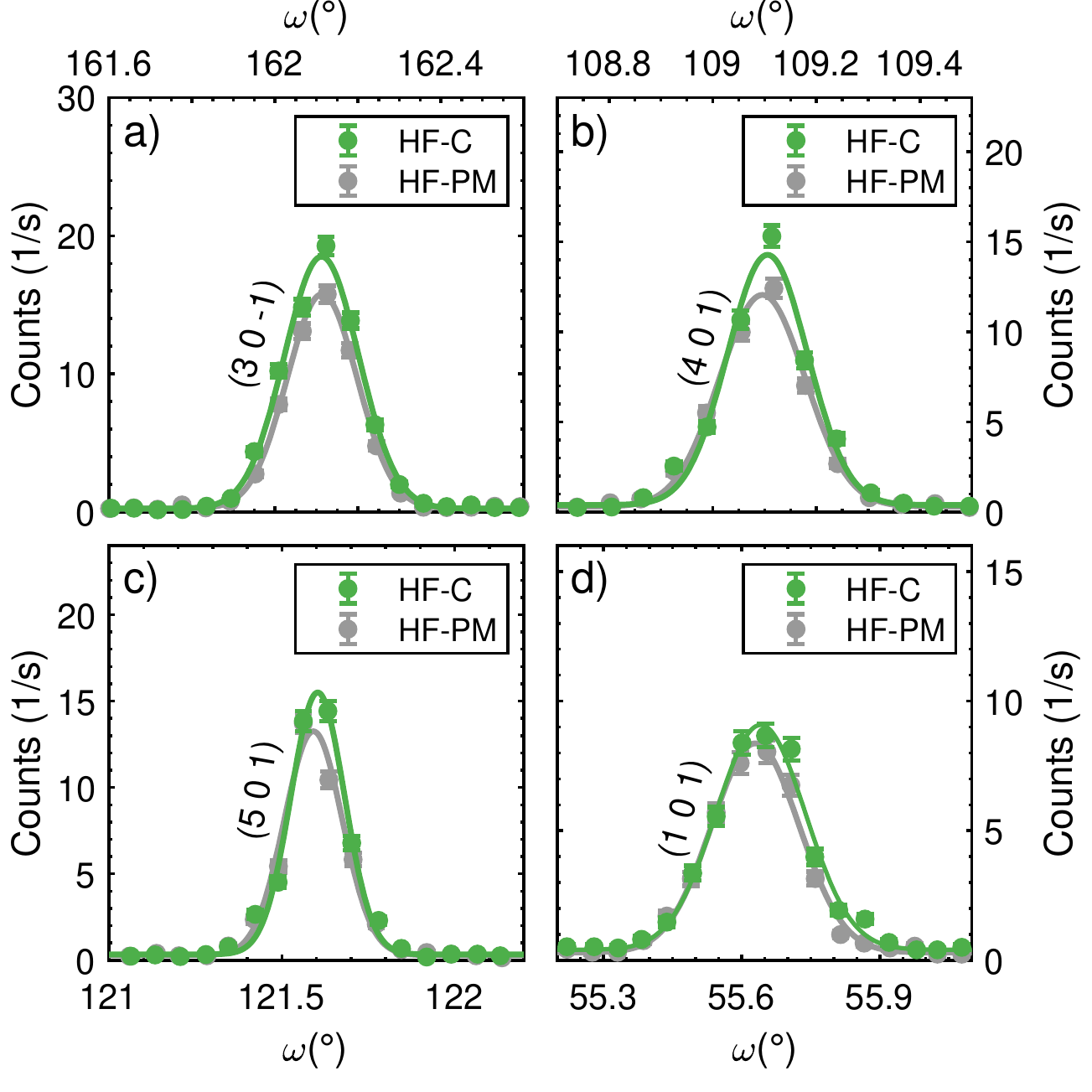}
  \caption{\label{rocking_scans_nafe}\nafekomma{:} (a) - (d) display rocking-scans over the $Q$-space position of superstructure reflections. For all measured reflections, the respective rocking scans have been executed in the high-field paramagnetic phase (HF-PM) at $T\approx\SI{10}{\kelvin}$ and $\mu_0H=\SI{6}{\tesla}$ as well as within the high-field commensurate phase (HF-C) at $T\approx\SI{2.4}{\kelvin}$ and $\mu_0H=\SI{6}{\tesla}$.}
  \label{disp}
 \end{figure}
 While decrasing the magnetic field again (below $\mu_0H\approx \SI{2}{\tesla}$ at $T\approx \SI{2.5}{\kelvin}$), a spin-flop transition forces a spin-canting but the structure remains commensurate and still exhibits an $uudd$ arrangement. This LF-C phase has been determined to be the groundstate of the system and all transitions from incommensurate to commensurate order are reported to be of first order \cite{Holbein2016}.

 Astonishingly, it is not the onset of magnetic ordering that is accompanied by strong magnetoelastic anomalies but rather the emergence of anharmonic modulations  and especially the transitions from incommensurate to the commensurate phases, which emphasizes a strong magnetoelastic coupling to them \cite{Holbein2016}.

 \begin{figure}
 \includegraphics[width=\columnwidth]{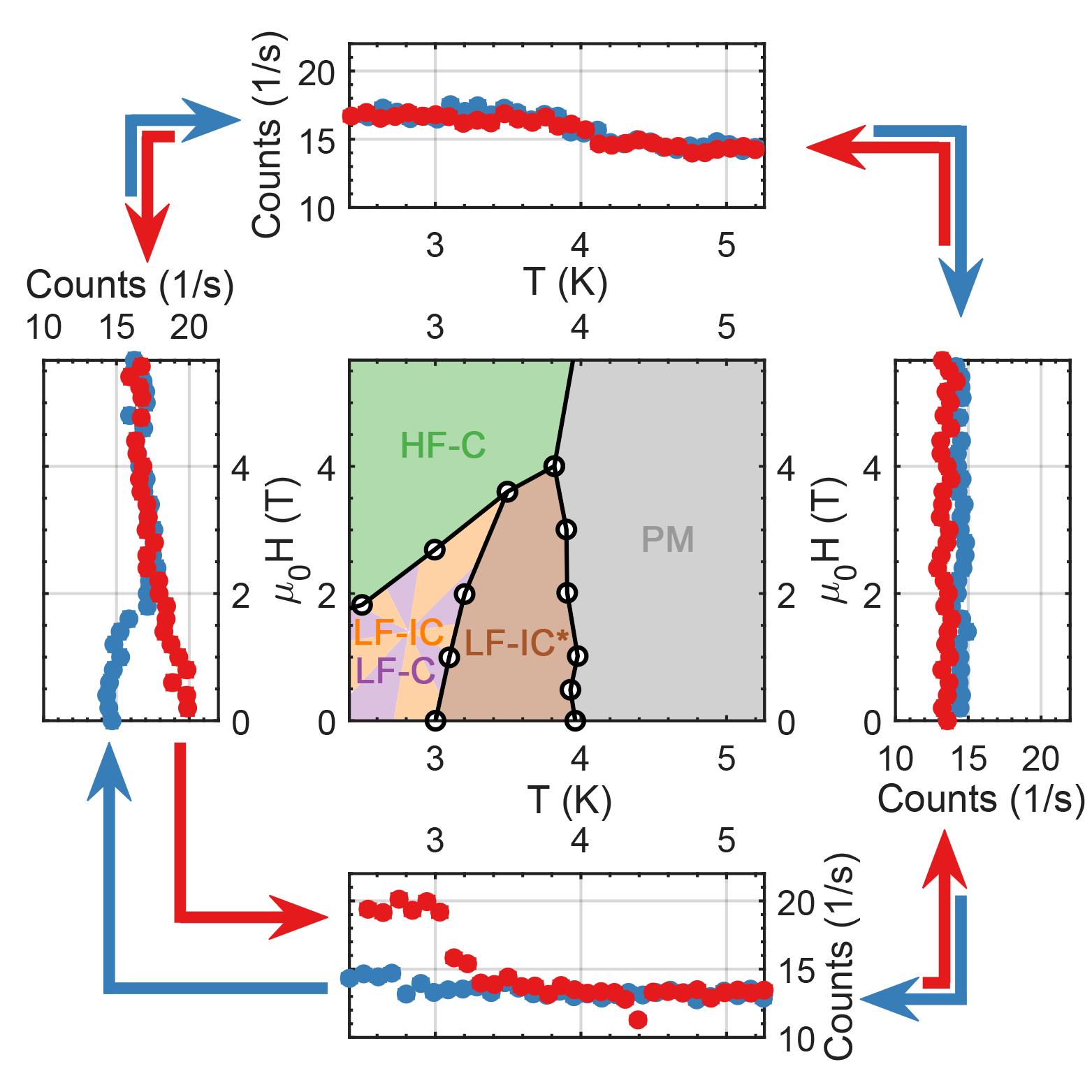}
  \caption{\label{sweep_1}\nafekomma{:} For analyzing the occurrence and strength of superstructure reflections in a particular phase, temperature and magnetic field dependent measurements have been performed exemplary for the superstructure reflection $Q=(3\:0\: \text{-}1)$. The temperature and magnetic field dependent sweep path is here defined by the axes of the phase diagram (the white marked phase boundaries are extracted from \cite{Holbein2016}), which is plotted in the middle of this figure. The sweep path was traversed clockwise (blue path) and also anticlockwise (red path), as the accessibility of the LF-IC and LF-C phase depends on pre-history of applied fields. The $T$- and $\mu_0H$-sweep speed was low, as the weak signal required a counting time of $\approx 3$ minutes per point.}
  \label{disp}
 \end{figure}

\begin{figure}
 \includegraphics[width=\columnwidth]{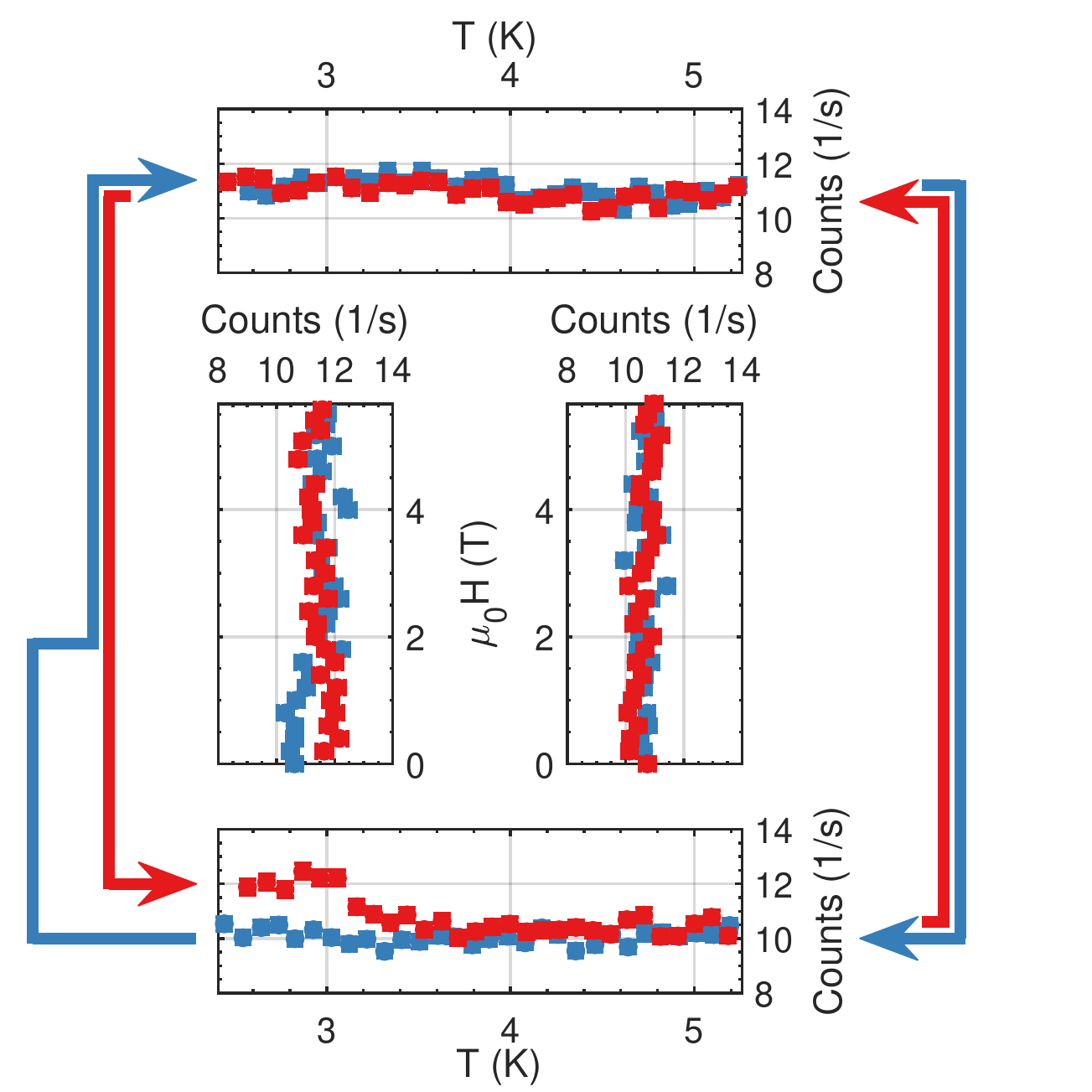}
  \caption{\label{sweep_2}\nafekomma{:} Similar to Fig.~\ref{sweep_1} this figure presents the temperature and magnetic field dependence of the the peak intensity for the superstructure reflection $Q=(4\:0\:1)$ in \nafe . The blue and red arrows indicate the clockwise or anticlockwise paths in the $B$,$T$ space respectively.}
  \label{disp}
 \end{figure}

The main goal of the neutron experiments was to determine the structural distortion within the commensurate phases of the system. A bond-angle variation along the Fe-chain due to alternating exchange interaction violates the glide plane and the twofold rotation symmetry, but it does not break translation symmetry. Therefore, superstructure reflections \cite{superstructure_reflection} will appear at $Q=(h\:0\:l)$ with $l=\text{odd}$. Fig.~\ref{rocking_scans_nafe} displays recorded rocking-scans across several superstructure reflections. All reflections have been measured  inside the HF-C phase at $T\approx\SI{2.4}{\kelvin}$ and within the paramagnetic phase at $T\approx\SI{5.2}{\kelvin}$, while applying a magnetic field of $\mu_0H=\SI{6}{\tesla}$.
Even in the paramagnetic phase a reflection is detectable, whose presence should actually be forbidden due to the $P2/c$ symmetry.
The field induced moment cannot contribute a magnetic signal at these $Q$ values, because it does not break the glide-mirror symmetry.
This is also visible in the magnetic field dependent measurements in Fig.~\ref{sweep_1} and \ref{sweep_2} for $Q=(3\:0\: \text{-}1)$ and $Q=(4\:0\: 1)$ respectively.
This Bragg-signal in the paramagnetic phase can arise from either $\lambda/2$-contamination or multiple scattering. The $\lambda/2$-contamination
was determined through rocking-scans across the allowed $(5\:0\:2)$ reflection and the resulting $\lambda/2$-contamination at $Q=(2.5\:0\:1)$.
The fraction of $\lambda/2$-contamination with respect to the main reflection amounts to $\approx 6\times 10^{-4}$ and is too small to explain the observed $Q=(3\:0\: \text{-}1)$ and $Q=(4\:0\: 1)$ intensities
in the paramagnetic state. By measuring the intensity at $Q=(6\:0\:\text{-}2)$, we can conclude that the $\lambda/2$-contamination at $Q=(3\:0\: \text{-}1)$  only approaches $\approx 1.32$ counts per seconds (see Fig.~\ref{rocking_scans_nafe}(a)).
Thus,  multiple scattering almost entirely causes the finite intensities at $Q=(3\:0\: \text{-}1)$ and $Q=(4\:0\: 1)$ outside the
commensurate phases, but the enhancement of these signals in the commensurate phases cannot stem from multiple diffraction \cite{note1}.
In Fig.~\ref{rocking_scans_nafe}, it can be seen that the integrated intensity of all reflections increases, when forcing the system to the HF-C phase. Therefore, we can conclude that the glide plane symmetry is broken, when entering the HF-C phase.

In order to investigate the occurrence and the strength of superstructure reflections in all phases, we performed temperature and magnetic-field dependent sweeps through the phase diagram. Due to the weakness of the signal and the corresponding long counting time, we restricted the data collection on recording only the intensity at the maximum position of the peak. Nevertheless, we checked by some rocking scans, whether the Bragg-peak shifts for different fields and temperatures. As this was not the case, we can exclude a maximum peak intensity modulation due to a shift of the Bragg-reflection in $Q$-space.  Exemplary for the $Q=(3\:0\: \text{-}1)$ superstructure reflection, the temperature and magnetic-field dependent intensity is shown in Fig.~\ref{sweep_1}. Because the commensurate or incommensurate character of the low-field phases strongly depends on the pre-history of applied magnetic fields in this system, we performed clockwise and anticlockwise sweeps through the phase diagram of \nafekomma{.} The temperature and magnetic field dependent sweep paths are marked by the axes of the $B$,$T$ phase diagram in the middle of Fig.~\ref{sweep_1}.

Starting the clockwise cycle (blue symbols in Fig.~\ref{sweep_1}) in the paramagnetic phase at $\mu_0H=\SI{0}{\tesla}$ and $T\approx\SI{5.2}{\kelvin}$, a reduction of the temperature does not change the peak intensity of the superstructure reflection, when passing the two transitions to the incommensurate phase (LF-IC$^*$) and to the anharmonic distorted incommensurate phase (LF-IC) respectively. Increasing the magnetic field at low temperatures leads to a sudden intensity enhancement at $\mu_0H\approx\SI{1.8}{\tesla}$. At that field value, the system undergoes a transition from an incommensurate magnetic ordering to a commensurate $uudd$ arrangement. The intensity enhancement and hence, the occurrence of a superstructure reflection coincides with the onset of an alternating ferro- and antiferromagnetic exchange along the magnetic zigzag chains. When increasing the temperature at high fields, the intensity drops back again to its initial value right at the transition to the paramagnetic phase at $T\approx\SI{4}{\kelvin}$ and lowering the field at
5.2\,K does not further change the intensity.

The anticlockwise sweep (red symbols in Fig.~\ref{sweep_1}) through the phase diagram recovers the measured intensity of the clockwise sweep, when starting again from the paramagnetic phase in zero field at $T\approx\SI{5.2}{\kelvin}$ ramping the field up and then cooling at high field. This is not surprising, as the same high-field phase is reached, when cooling below $T\approx\SI{4}{\kelvin}$ in high fields at $\mu_0H=\SI{6}{\tesla}$. However, when decreasing the field at low temperatures, the magnetic arrangement does not return to the incommensurate and metastable low-field phase but a spin-flop transition at $\mu_0H\approx\SI{1.8}{\tesla}$ leads to the LF-C phase with canted moments that are still ordered in a commensurate $uudd$ arrangement. The superstructure intensity even further increases, when removing the applied field completely and hence, the superstructure reflection is strongest for the groundstate of the system, which leads to the assumption that also the bond-angle variation is strongest for this LF-C phase.
The intensity sharply decreases, when the temperature is raised above the transition to the incommensurate LF-IC* phase recovering its initial value, which limits the presence of the structural distortion precisely to the commensurate phases. Qualitatively identical behavior is observed for the superstructure reflection $Q=(4\:0\:1)$ (see Fig.~\ref{sweep_2}).

\subsection{Quantitative determination of the dimerization in \nafe }

For a structural refinement and hence the quantitative analysis of the structural distortion, we collected a set of integrated intensities in the dimerized phase on the D10
diffractometer.  As the superstructure signal is strongest within the LF-C phase, see Fig.~\ref{sweep_1} and \ref{sweep_2}, a structural refinement seems to be most promising to execute within this commensurate phase in zero field. Moreover the data collection in zero field avoids an additional signal contribution from field induced moments to allowed main reflections. However, the application of a magnetic field was still necessary in order to drive the system to the LF-C phase.

Before we collected respective superstructure reflections, we measured (with a microstrip detector and without additional analyzer) 108 unique allowed nuclear reflections of which 56 are independent within the LF-IC phase in order to refine the isotropic extinction coefficient. The refinement was done by using the software JANA \cite{Petricek2014}, which utilizes the Becker-Coppens formalism for the extinction correction. For the corresponding refinement  we used the structural parameters that have been obtained by Holbein $et$ $al.$ \cite{Holbein2016}. Hereafter we measured with a single He$^3$-detector and with the analyzer option to improve the intensity over background ratio. In order to obtain the appropriate scaling factor for the refinement of the superstructure, we measured 13 unique allowed reflections, of which 7 are independent at $T\approx\SI{2}{\kelvin}$ and $\mu_0H=\SI{0}{\tesla}$.
\begin{table*}
\caption{\label{tab:table3}The left part of this table contains the structural parameters that have been determined in the paramagnetic phase at $T=\SI{12}{\kelvin}$ by Holbein $et$ $al.$ \cite{Holbein2016}. All positions have be converted to space group $P\bar{1}$ as given in the middle part of this table. The right part consists of the refined structural parameters within the LF-C phase and by assuming $P\bar{1}$ symmetry. For the refinement 26 unique superstructure reflection of which 24 are independent have been used. Moreover 13 unique allowed reflections of which 7 are independent have been added to the refinement. In order to reduce the number of variables, the following constraints have been utilized: x[O2a]=x[O2b]-0.78224, x[W1]=x[W2]-0.52592, z[O2a]=z[O2b]-0.3154 and z[W1]=z[W2]+0.0144. The structural refinement yielded reliability factors of $R_{\text{obs}}=7.07$, $wR_{\text{obs}}=16.44$, $R_{\text{all}}=9.13$ and $wR_{\text{all}}=18.54$ for structure factors.}
\begin{ruledtabular}
\begin{tabular}{cccccccccccc}
 & \multicolumn{4}{c}{Paramagnetic phase (spacegroup $P2/c$)}     &  & \multicolumn{3}{c}{Paramagnetic phase (spacegroup $P\bar{1}$)}   & \multicolumn{3}{c}{LF-C phase (spacegroup $P\bar{1}$)} \\

 &x  &y  & z &B$_\text{iso}$  &  &x  &y  &z  & x&y&z \\ \hline

 Fe&0.0  &0.67074(19)  &0.25  &0.04(2)  &Fe  & 0.0 & 0.67074(19) & 0.25 &-0.0002(2) &0.67074& 0.2525(7) \\

 Na&0.5  &0.6971(6)  &0.25  &0.35(5)  &Na  & 0.5 & 0.6971(6) & 0.25 &0.5  &0.6971&0.25\\

 W&0.23704(14) &0.1831(2)  &0.2572(3)  &0.12(3)  &W1  &0.23704(14)  & 0.1831(2) &0.2572(3)  & 0.2383(3) & 0.1831 &0.2562(8) \\

 O1&0.35385(12)  &0.3813(3)  &0.3816(3)  &0.25(3)  &W2  & 0.76296(14) & 0.1831(2) & 0.2428(3) &0.7642(3) &0.1831& 0.2418(8)\\

 O2&0.10888(13)  &0.6226(3)  &0.5923(3)  &0.22(2)  &O1a  & 0.35385(12) & 0.3813(3) & 0.3816(3) &0.35385 &0.3813&0.3816 \\

 O3&0.33177(13) &0.0897(2)  &0.9533(3)  &0.22(3)  &O1b  & 0.64615(12) & 0.3813(3) & 0.1184(3) &0.64615 &0.3813& 0.1184\\

 O4&0.12606(13) &0.1215(3)  & 0.5757(3) &0.17(2)  &O2a & 0.10888(13) & 0.6226(3) & 0.5923(3) & 0.1111(2)&0.6226&0.5935(8) \\

 &  &  &  &  & O2b & 0.89112(13) & 0.6226(3) & 0.9077(3) &0.8933(2) &0.6226&0.9089(8) \\

 &  &  &  &  & O3a  &0.33177(13)  & 0.0897(2) & 0.9533(3) &0.33177 &0.0897&0.9533 \\
 &  &  &  &  & O3b  &0.66823(13)  & 0.0897(2) & 0.5467(3) &0.66823 &0.0897&0.5467 \\

 &  &  &  &  & O4a & 0.12606(13) & 0.1215(3) & 0.5757(3) &0.12606  &0.1215&0.5757\\

 &  &  &  &  & O4b  & 0.87394(13) & 0.1215(3) & 0.9243(3) &0.87394 &0.1215& 0.9243
\end{tabular}
\end{ruledtabular}
\end{table*}

It was already discussed that the intensity of  superstructure peaks superposes the  multiple scattering signal and the tiny $\lambda/2$-contamination and thus, it was necessary to measure the respective superstructure reflections inside and outside the commensurate phases in order to extract the integrated intensity that is purely originating from the structural distortion. Hence, we measured these superstructure reflections in the LF-C phase at $T\approx\SI{2}{\kelvin}$, $\mu_0H=\SI{0}{\tesla}$ as well as in the LF-IC phase (also at $T\approx\SI{2}{\kelvin}$, $\mu_0H=\SI{0}{\tesla}$). This procedure and the weak signal itself limit the number of collectible reflections and we were able to collect 26 superstructure reflections of which 24 are independent. The required vertical cryomagnet with its magnetic field along $b$  and the analyzer option prevented the collection of reflections with a non-zero $k$-component and thus the structural refinement will only testify a structural distortion within the $ac$-plane.

The left part of Table \ref{tab:table3} contains the structural parameters that have been refined by Holbein $et$ $al.$ in the paramagnetic phase \cite{Holbein2016}. All Fe ions are surrounded by edge sharing oxygen octahedra, and when considering the $P2/c$ space group, all oxygen ions along the Fe-chain are generated from the general site O2.
The distance between two O2's belonging to the same edge alternate in the dimerized phase resulting in alternating enhancement and reduction of the Fe-O2-Fe bond angle
and thus in an alternation of the magnetic exchange energy.
In the simple picture neglecting ligands, a ferromagnetic arrangement prefers a $\SI{90}{\degree}$ Fe-O-Fe bond-angle, whereas an antiferromagnetic bond requires the bond angle to approach $\SI{180}{\degree}$.
Both O2 ions on the same edge are connected through inversion symmetry, and as they elongate or shorten their distance uniformly, the inversion symmetry is preserved in a dimerized state. However, an alternating exchange interaction along the chain and the corresponding bond-angle modulation violate the two-fold axis and the glide mirror symmetry
giving rise to the observed superstructure reflections. Thus, the refinement of the superstructure within the LF-C phase was done by considering a transition from space group $P2/c$ to $P\bar{1}$. The middle part of Table \ref{tab:table3} displays the structural parameters transformed from $P2/c$ to $P\bar{1}$ setting and the right part the refined distortion
in the lower symmetry.

\begin{figure}
 \includegraphics[width=\columnwidth]{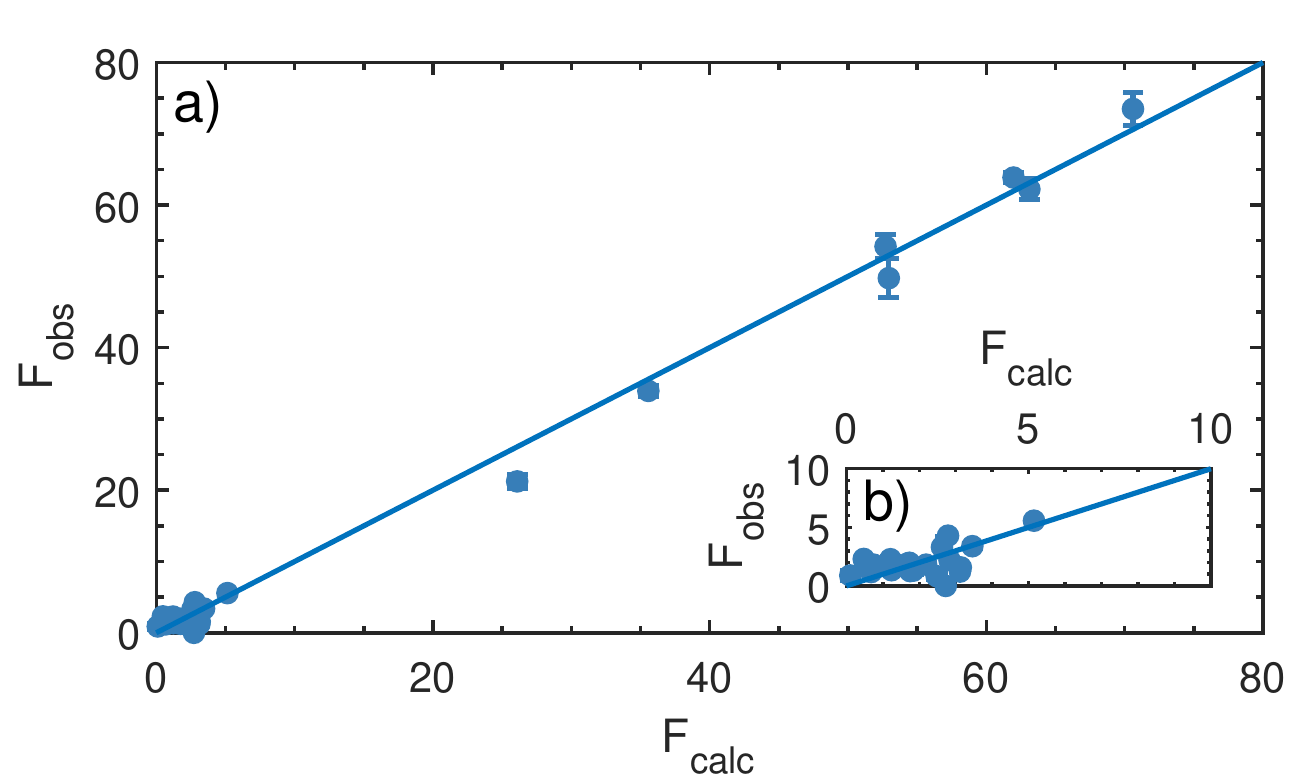}
  \caption{\label{plot-i-i} The observed structure factors are plotted against the calculated ones for the final refinement of the structural dimerization within the LF-C phase. }
  \label{disp}
 \end{figure}

The transformation to space group $P\bar{1}$ splits the site of O2 in P2/c into the two independent sites O2a and O2b. The Fe-O-Fe bonding along the chains passes alternatingly through
O2a-O2a and O2b-O2b edges. In the refinements we kept the average O2 position fixed and allowed only for alternating O2 shifts.
In order to further reduce the number of variables, equivalent constraints have been also stated for the W1 and W2 positions. 6 variables have been refined. The right part of Table \ref{tab:table3} contains the refined parameters of the $P\bar{1}$ structure within the LF-C phase. For the refinement, the isotropic B parameters have been adopted from \cite{Holbein2016}. At first glance, the obtained R-values (see Table \ref{tab:table3}) may appear large but one has to keep in mind that only very weak superstructure reflections (almost 4 orders of magnitude weaker than
strong fundamental Bragg peaks) intervene in these refinements.
In Fig.~\ref{plot-i-i} we plot the observed structure factors against the calculated ones documenting the reliability of this analysis.
The O2a and O2b sites are shifting significantly in opposite directions by $\SI{0.0195(32)}{\angstrom}$. The Fe ions are only moving along $z$ by $\SI{0.0093(31)}{\angstrom}$ and the position of W1 and W2 is also shifting because of their bonding to the O2a and O2b sites. Within the commensurate LF-C phase, the calculated bond angles amount $\alpha_1=\SI{97.20(14)}{\degree}$ for the Fe-O2a-Fe bond and $\alpha_2=\SI{99.49(14)}{\degree}$ for the Fe-O2b-Fe bond, whereas the bond angle outside the commensurate phase amounts $\alpha=\alpha_1=\alpha_2=\SI{98.35(7)}{\degree}$ \cite{Holbein2016}. Hence, a bond-angle variation of $\approx\pm\SI{1.15(16)}{\degree}$ can be observed, when entering the commensurate $uudd$ phase. This modulation includes the shift of oxygen and iron ions within the $ac$-plane.
A respective shift along $b$ cannot be studied with the chosen scattering geometry.

The implication of the bond angle, $\phi$, on the magnetic exchange interaction has been intensively studied for $RMe$O$_3$ with $Me$=Fe and Mn finding a $\sin^4(\phi/2)$ relation \cite{Zhou2008,Dong2008}.
This strong bond-angle dependence of the magnetic interaction drives the magnetoelectric coupling in E-type $R$MnO$_3$ \cite{Sergienko2006,Ishiwata2010} and causes the dynamic magnetoelectric coupling and the strongest electromagnon modes in chiral $R$MnO$_3$ \cite{ValdesAguilar2009}.
However, this $\sin^4(\phi/2)$ relation applies to a bond angle, which is not too far
below 180{\deg}, while the O-Fe-O angles in \nafe \ amount to 97.2 and 99.5 degrees.
Therefore, the arrangement more closely resembles the dimerization in the spin-Peierls compound
CuGeO$_3$, in which a bond-angle alternation of  only $\pm0.5$\deg \  around the average value of $\sim$99\deg \ results in a sizeable variation of the magnetic exchange by $\pm$20 percent \cite{Braden1996}.
Unfortunately, the magnetic exchange parameters in \nafe \ are not known. For MnWO$_4$ the nearest-neighbor intra-chain is not dominating and several parameters are of the order of 0.3 to 0.4\,meV \cite{Ye2011,Tian2009} explaining the considerable frustration.
The antiferromagnetic Curie-Weiss temperature in \nafe \ amounts to -8.5\,K (compared to -75\,K for MnWO$_4$ \cite{Tian2009}), which  transforms to a nearest-neighbor intrachain interaction of only 0.06\,meV when neglecting all other interactions.  \nafe \ seems to be less frustrated with a ratio of N\'eel and Weiss temperatures of only $\sim$2, but the next-nearest neighbor interaction along the chains will stabilize the $uudd$ magnetic order. Since the magnetic interaction in \nafe \ is quite weak the relative change induced by the sizeable structural dimerization through the bond-angle alternation is expected to be rather strong.

The strong reduction of the $b$ lattice parameter upon entering the commensurate magnetic order in \nafekomma{} \cite{Holbein2016} 
is not contributing to the dimerization, which can be considered
as the primary order parameter of the structural distortion. Note, that the  strain does not break the glide mirror  symmetry. 
Since both distortions correspond to zone-center phonon modes, one
would expect a bilinear coupling, but the coupling with the magnetism renders the analysis more complex. 
The commensurate magnetic order clearly is the the primary order
parameter and both structural distortions, dimerization and strain along $b$, are coupled to it. 
The dimerization modulates the nearest-neighbor interaction, thereby rendering the $uudd$ structure more favorable. The strain along $b$ acts on the ratio
between nearest and next-nearest neighbor interaction. One expects an enhanced second-nearest neighbor interaction for smaller $b$, which again
will favor the commensurate $uudd$ order. Since the two types of distortion act differently, one can understand their different relative strengths
in \nafekomma{} and MnWO$_4$.

\subsection{\label{sec:level2}Structural dimerization in \mnwo}

\mnwo crystallizes in the monoclinic space group $P2/c$ with $a\approx\SI{4.823}{\angstrom}$, $b\approx\SI{5.753}{\angstrom}$, $c\approx\SI{4.992}{\angstrom}$ and $\beta\approx\SI{91.08}{\degree}$ (at $\SI{1.5}{\kelvin}$) \cite{Lautenschlager1993}. Within the $bc$-plane, magnetic $\text{Mn}^{2+}$ ions are forming edge-sharing [MnO$_6$] octahedra chains along $c$ and those layers, containing the chains are separated along the $a$ direction by nonmagnetic [WO$_6$]-octahedra chains \cite{weitzel1976}. The crystallographic unit cell of the parent wolframite structure is bisected along $a$ with respect to the structure of \nafe due to the missing [NaO$_6$]-layer. Similar to \nafekomma{,} the system exhibits also a sequence of magnetic phases with competing commensurate and incommensurate ordering \cite{Lautenschlager1993,Ehrenberg1997a}. First, below $T_{N3}\approx\SI{13.5}{\kelvin}$ an incommensurate modulated spin-density wave with $\bm{k}=(\text{-}0.214,\frac{1}{2},0.457)$ emerges (AF3-phase). The magnetic moments point along the easy direction, which lies within the $ac$-plane and forms an angle of about $\approx\SI{35.5}{\degree}$ with the $a$-direction. In this configuration, the magnetic moments cannot saturate and thus due to entropy reasons a spiral with an additional component along the $b$-direction develops at $T_{N2}\approx\SI{12.3}{\kelvin}$ but the modulation remains incommensurate (AF2-phase)\cite{Lautenschlager1993,Ehrenberg1997a}. The spiral arrangements allows for the occurrence of a ferroelectric polarization following DMI and indeed, it was possible to measure an electric polarization  along $b$, which is fully switchable by external electric fields \cite{Niermann2014,Baum2014}. The polarization is small but sizeable ($P\approx\SI{60}{\micro\coulomb\per\square\meter}$) and its existence is limited to the spiral phase AF2 \cite{Arkenbout2006,Taniguchi2006}. However, due to single ion anisotropy the system preferably aligns along its easy direction, which provokes the transition to a commensurate $uudd$ arrangement with $\bm{k}=(\text{-}\frac{1}{4},\frac{1}{2},\frac{1}{2})$ at $T_{N1}\approx\SI{7.5}{\kelvin}$\cite{Lautenschlager1993,Ehrenberg1997a}.

As in \nafekomma{,} the transition to the commensurate phase is of first order and accompanied by a thermal expansion anomaly. It is expected that a structural distortion is here also caused by a bond-angle modulation due to the alternating ferro- and antiferromagnetic exchange interaction along the magnetic Mn$^{2+}$ chain. The observed thermal expansion anomaly in \mnwo is one order of magnitude smaller compared to \nafe but its commensurate phase is accessible without the application of a magnetic field. The symmetry consideration from \nafe can be adopted to \mnwo and it becomes obvious  that corresponding superstructure reflections occur at $Q=(h$+$\frac{1}{2}\:0\:l)$ with $l$=odd for \mnwokomma{.}  X-ray measurements have already qualitatively confirmed the presence of a respective signal in the context of second harmonic order reflections within the AF1 phase \cite{Taniguchi2008}.

\begin{figure}
 \includegraphics[width=\columnwidth]{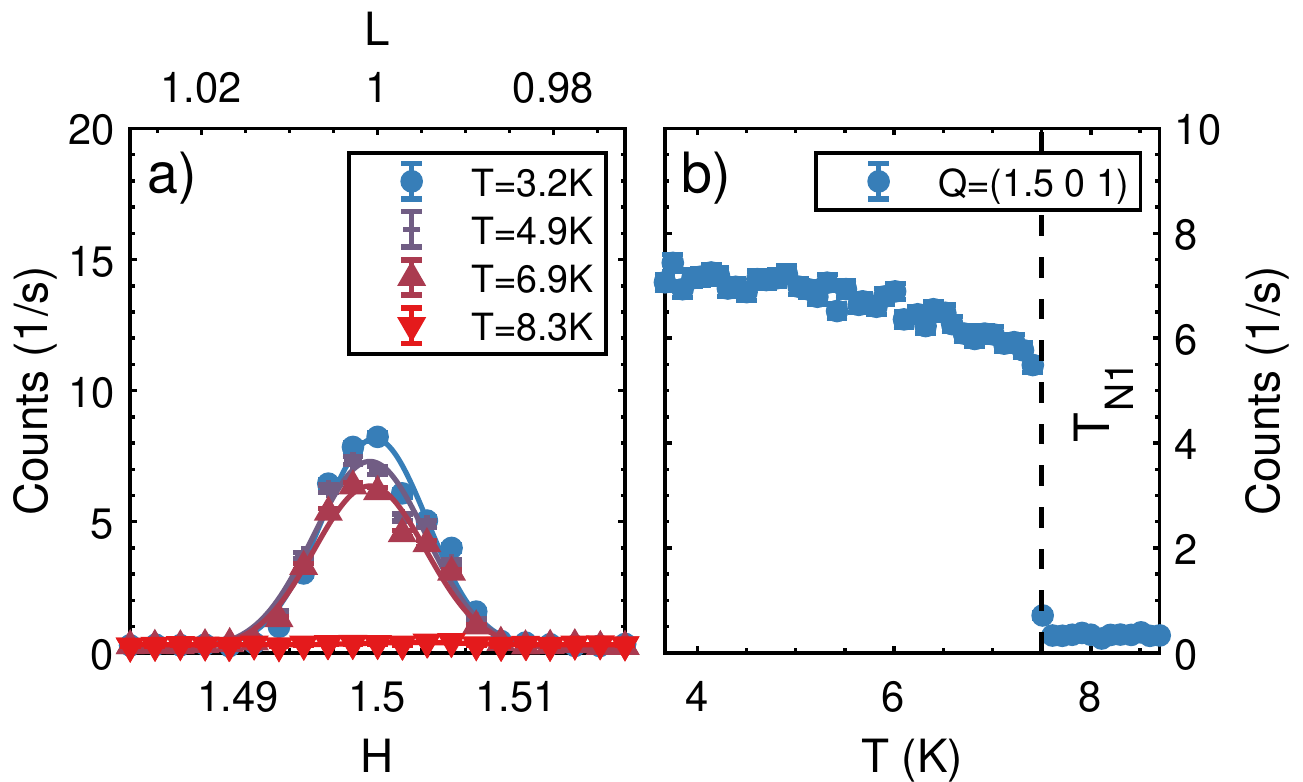}
  \caption{\label{dimerization_mnwo}\mnwo{:} Panel (a) shows the longitudinal scans over $Q=(H\:0\:L)$ for four different temperatures. The measured peak intensity for $Q=(1.5\:0\:1)$ is displayed in panel (b) as a function of temperature.}
  \label{disp}
 \end{figure}

Neutron scattering experiments were performed on the cold triple-axis spectrometer 4F2, where we first executed longitudinal scans over the Q-space position of the superstructure reflection $Q=(1.5\:0\: 1)$ for different temperatures. At these noninteger positions no multiple scattering can contribute to the signal. From the measured data, which is displayed in Fig.~\ref{dimerization_mnwo}(a) it can be clearly stated that at the transition to the low-lying commensurate phase, a superstructure reflection evolves. We measured the peak intensity, while lowering the temperature from above $T_{N1}$ to far below the transition of the commensurate phase (see Fig.~\ref{dimerization_mnwo}(b)). This temperature dependent sweep demonstrates that the superstructure reflection is exactly evolving, when passing the latter mentioned transition.
Hence, also in \mnwo the occurrence of the commensurate $uudd$ arrangement coincides with an evolving superstructure.
In view of the equivalent behavior in \nafekomma{,} we conclude that also in \mnwo a bond-angle modulation is coupled to the onset of commensurate $uudd$ ordering. Comparing the intensities of superstructure reflections $Q=(2h\:0\:l)$ for \nafe and $Q=(h\:0\:l)$ for \mnwo ($h$=1.5) with respect to the particular strongest allowed reflections ($Q=(0\:2\:3)$\cite{note2} in \nafe and $Q=(1\:0\:0)$ in \mnwokomma{)} results in almost equal ratios of a few $10^{-4}$. This indicates a bond-angle modulation of the same order of magnitude in the two systems.

Locally, the commensurate $uudd$ structure can also evolve within the multiferroic phase and it was proposed for \nafe that an anharmonic modulation of the incommensurate structure within its incommensurate LF-IC phase corresponds to tiny fragments of commensurate order \cite{Holbein2016} that are regularly arranged. This assumption is supported by the occurrence of third order reflections, which signal a squaring up of the cycloid \cite{Holbein2016}.

\begin{figure}
 \includegraphics[width=\columnwidth]{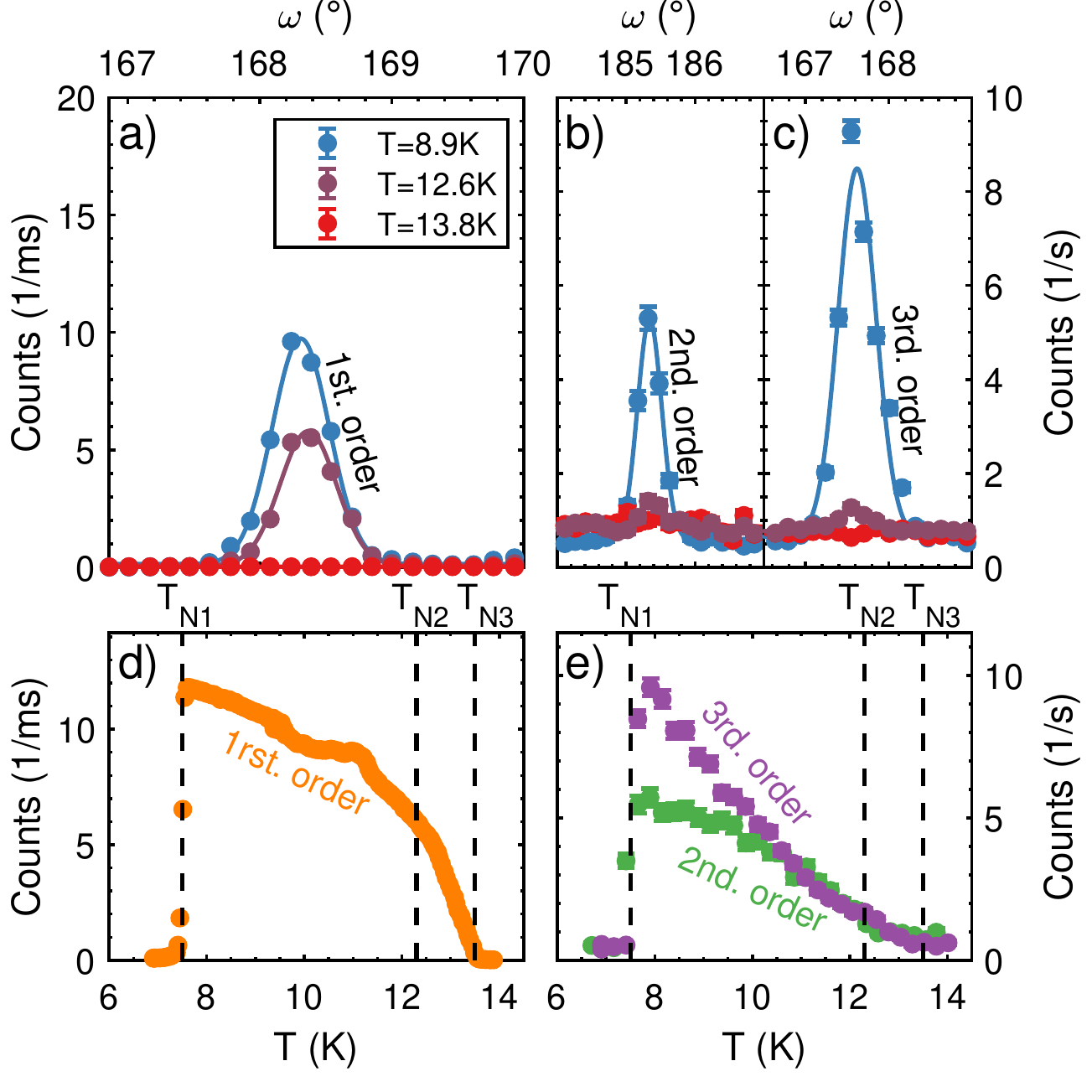}
  \caption{\label{higher_harmonics}\mnwo{:} Panels (a), (b) and (c) present rocking scans over the first ($Q=($-$0.214\:\frac{1}{2}\:0.457)$), second ($Q=($-$0.428\:1\:0.914)$) and third harmonic reflection ($Q=($-$0.642\:\frac{1}{2}\:1.371)$) for different temperatures. The respective temperature dependence of the peak intensities is shown in (d) and (e).}
  \label{disp}
 \end{figure}

Fig.~\ref{higher_harmonics}(a)-(c) presents rocking scans of the higher harmonic reflections as a function temperature for \mnwokomma{.} In Fig.~\ref{higher_harmonics}(d) and (e), the peak intensity is plotted as a function of temperature. The temperature dependence of the first and of the second order reflections resembles reported  measurements \cite{Finger2010} but the occurrence of a third order reflection has not been reported so far. Fig.~\ref{higher_harmonics}(e) displays the temperature dependence of the measured peak intensity of the third order reflection $Q=(\text{-}0.642\:\frac{1}{2}\:1.371)$. At low temperatures this reflection becomes even stronger than the  second-order reflection and its existence is not limited to the multiferroic phase but a finite signal persists in the AF3 phase.
While the second-order intensity clearly follows the intensity of the main magnetic peak, the third-order reflection exhibits a different temperature dependence.
The first- and second-order peaks tend to saturate towards the lower temperature limit of the multiferroic AF2 phase, but the growth of the third-order peak
continuously increases showing a negative second temperature derivative. This strongly resembles the behavior in \nafe \ \cite{Holbein2016} and this diverging
third-order peak must be considered as a precursor of the phase transition into the commensurate $uudd$ phase. Tiny commensurate fragments with
the corresponding discommensurations strongly increase upon cooling towards the low-temperature limit of the AF2 phase. These perturbations of
the harmonic multiferroic magnetic structure can explain the depinning of multiferroic ordering reported in the various studies on the the
dynamics of multiferroic domains \cite{Niermann2014,Baum2014,note-average}.

 \section{\label{sec:level1}Summary and Conclusion}

We present a comprehensive study of a dimerization effect in \nafe and \mnwokomma{.} It was possible to qualitatively observe the existence of superstructure reflections documenting the loss of symmetry, when entering the respective commensurate magnetic phases of both materials.
This loss of symmetry is connected to a bond-angle modulation along the magnetic zigzag chains due to the alternating ferromagnetic and antiferromagnetic arrangement in the respective commensurate $uudd$ phases.
For \nafe a bond-angle modulation is present in the HF-C as well as in the LF-C phase and it turned out that the dimerization effect is strongest for the commensurate phase in zero field.
In this phase, a structural refinement of neutron data yields a bond-angle modulation of $\approx\pm\SI{1.15(16)}{\degree}$ but it has to be noted that this specified bond-angle modulation doesn't include a possible modulation along the $b$-direction, as the chosen scattering plane and the utilized cryomagnet and analyzer prevented the collection of reflections with a finite $k$-component.

Also for \mnwokomma{,} a structural dimerization of comparable size was detected by the appearance of superstructure reflections in its commensurate phase.
Moreover, strong harmonic third order reflections have been observed within the multiferroic phase that perfectly correspond to the
anharmonic modulation reported for \nafe \cite{Holbein2016}. The emerging third-order reflection is connected to a squaring up of the magnetic ordering and indicates the presence of small commensurate fragments already in the incommensurate and multiferroic phase. As the magnetic arrangement of \mnwo already starts to resemble the $uudd$ arrangement
within the incommensurate spiral phase, it can lead to depinning of the multiferroic domains and of the corresponding domain walls.
The pinning of the multiferroic domains bases on the harmonic modulation, which causes the ferroelectric polarization through the inverse DMI. Therefore, the anomalous speeding up of the multiferroic domain inversion can be attributed to the emergence of the anharmonic modulations. Such effects can even be beneficial for the application of multiferroics.

Single-crystal neutron diffraction data from the D10 diffractometer are available \cite{dataD10}.

\section{\label{sec:level1}Acknowledgements}


This work was funded by the Deutsche Forschungsgemeinschaft (DFG,
German Research Foundation) - Project number 277146847 - CRC 1238, projects A02 and B04.

\nocite{apsrev41Control}
\bibliographystyle{apsrev4-1}

\begin{thebibliography}{43}
\bibitem [{sup()}]{superstructure_reflection}%
  \BibitemOpen
  \href@noop {} {}\bibinfo {note} {{A superstructure is commonly connected to a
  loss of translational symmetry while here only a glide-mirror plane is
  broken. However, we will use the notation of a superstructure reflection also
  for the reflections $(h\:0\:l)$ with $l$ odd that are systematically absent
  in space group $P2/c$ due to the $c$-glide plane.}}\BibitemShut {Stop}%
\bibitem [{not()}]{note1}%
  \BibitemOpen
  \href@noop {} {}\bibinfo {note} {{We note that due to the usage of the
  vertical cryomagnet, it was not possible to modify the signal strength of
  multiple scattering by Renninger-scans.}}\BibitemShut {Stop}%
\bibitem [{not()}]{note2}%
  \BibitemOpen
  \href@noop {} {}\bibinfo {note} {{The reflection $Q=\left(0\:2\:3\right)$ was not measurable due to the chosen scattering geometry but however, the respective intensity can be calculated by comparing absolute square values of structure factors with the intensity of both, measured and non-measured reflections.}}\BibitemShut {Stop}%
\bibitem{note-average}{Note, that the speeding up of domain inversion when approaching the
 commensurate phases was also observed on large crystals presenting an average of domains.}
\end{thebibliography}
%

\end{document}